# Health Behavior Change in HCI: Trends, Patterns, and Opportunities

Yunlong Wang
HCI Group
University of Konstanz
Konstanz, Germany
yunlong.wang@uni-kosntanz.de

Ahmed Fadhil
ICT4G Group
Fondazione Bruno Kessler
Trento, Italy
fadhil@fbk.eu

Harald Reiterer
HCI Group
University of Konstanz
Konstanz, Germany
harald.reiterer@uni-konstanz.de

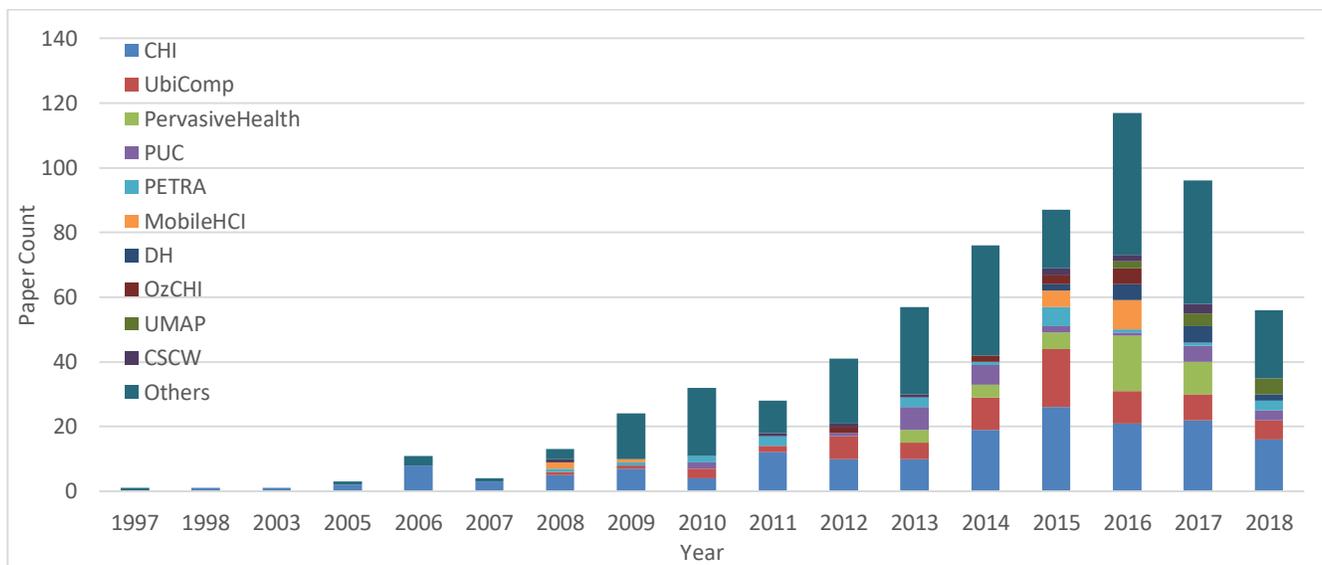

Figure 1: The paper frequency distribution in the field of health behavior change in the HCI community. Note that we extracted the original data from the ACM digital library on August 23, 2018, when some conferences for this year have not taken place. The majority of the papers are from conference proceedings, while a small part of them are from journals (e.g., Personal and Ubiquitous Computing or PUC in the figure).

## ABSTRACT

Unhealthy lifestyles could cause many chronic diseases, which bring patients and their families much burden. Research has shown the potential of digital technologies for supporting health behavior change to help us prevent these chronic diseases. The HCI community has contributed to the research on health behavior change for more than a decade. In this paper, we aim to explore the research trends and patterns of health behavior change in HCI. Our systematic review showed that physical activity drew much more attention than other behaviors. Most of the participants in the reviewed studies were adults, while children and the elderly were much less addressed. Also, we found there is a lack of standardized approaches to evaluating the user experience of interventions for health behavior change in HCI. Based on the reviewed studies, we provide suggestions and research opportunities on six topics, e.g., game integration, social support, and relevant AI application.

## 1 Introduction

Our daily behaviors heavily influence our health. According to the County Health Rankings [43], variation in health can be accounted for by health behaviors (30%), clinical care (20%), social and economic factors (40%), and physical environment (10%). Increasing evidence has shown that unhealthy behaviors - such as the unbalanced diet, inadequate physical activity, sleeping deprivation, drinking alcohol, and smoking - play an important role in individuals' health. Chronic diseases caused by unhealthy behaviors and habits are among the leading causes of mortality [18]. Some of the chronic diseases, e.g., type 2 diabetes, could be life-long and bring a heavy burden to the patients and their families. The only way to prevent many chronic diseases is to change unhealthy behaviors in the long term.

The research on digital technologies to support health behavior change is no doubt a vital task for the Human-Computer Interaction (HCI) community. Only in the proceedings of the ACM CHI conference until 2018, we found 310 papers mentioning "behavior change." However, it seems that the interest in health behavior change from the HCI community began to decrease recently. We see this trend by searching and screening the related papers from the ACM digital library. The amounts of the related papers from both the CHI conference and the UbiComp conference have seen the decrease since 2016, and the corresponding paper amount in CHI 2018 has fallen back to the level in 2014 (see Figure 1). To get an insight into this phenomenon, we conducted a systematic review of the papers about health behavior change from the HCI community. We extracted information from the perspectives of the target behavior, the target user group, the used behavioral theories, the deployed behavior change strategies, the intervention characteristics, and evaluation methods.

The remainder of this paper is organized as follows: The next section introduces behavioral theories, behavior change strategies, and behavioral intervention characteristics as the apparatus of our review. In Sect. 3, we show our methods to search, select, and code the studies. Sect. 4 reports our findings on research trends and patterns of health behavior change in HCI. Based on our reviewed studies, in Sect. 5, we provide suggestions and opportunities for the future research in six topics. Finally, we show the limitations of our work and conclude the paper.

## 2 Background

### 2.1 Behavioral Theories

Behavioral theories refer to the social-psychological theories of behavior change, which explain and predict human behavior. Glanz et al. [34] listed the most frequently used behavioral theories published before 2010: the Social Cognitive Theory (SCT) [3], the Transtheoretical Model of Change (TTM) [76], the Health Belief Model (HBM) [79], and the Theory of Planned Behavior (TPB) [2]. As explained by Sutton [80], each of the behavioral theories specifies a small number of cognitive and affective factors as the proximal determinants of behavior.

In a CHI paper in 2013, Hekler and colleagues illustrated the (dis)advantages of behavioral theories and how HCI researchers can use and contribute to behavioral theories [38]. In summary, behavioral theories can help inform design, guide evaluation strategies, and select target users. Also, HCI researchers have the change to improve behavioral theories by improving measurement, enhancing early-stage theory fidelity testing, and supporting and using big data and A/B testing. Following this implication, we will reveal how HCI research engaged with behavioral theories in the real world.

### 2.2 Behavior Change Techniques (Strategies)

Behavior change techniques (BCTs) are defined as observable, replicable, and irreducible components of an intervention designed to change behavior [1,59], e.g., self-monitoring of behavior and goal setting. Abraham and Michie published the taxonomy containing 93 BCTs in 16 groups in 2013, called Behavior Change Technique Taxonomy (v1) [59]. The BCT taxonomy has been used for informing intervention development [64,65] and identifying the effective ingredients in intervention studies for health behavior change [24,33,58,71] and products (i.e., health-oriented apps [17,20,23,60] and wearables [54]). The word cloud in Figure 2 shows the relative use frequencies of BCTs used in 405 studies. In the HCI community, however, BCT taxonomy is not used as widely as the model of Persuasive Technology [29] or Persuasive System Design (PSD) [70]. The model of PSD includes 28 principles in four categories, namely primary task support, dialogue support, system credibility support, and social support.

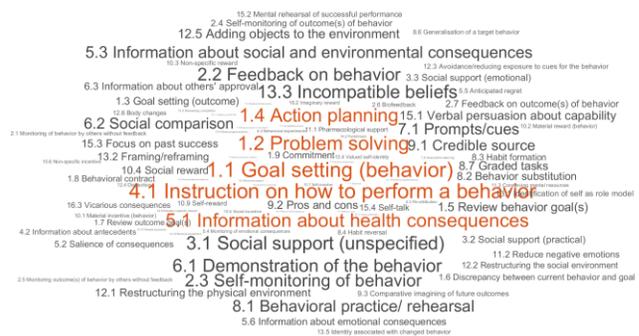

**Figure 2: The word cloud of behavior change strategies coded from the papers listed on the official website[1] of BCT taxonomy (N=405). The bigger the font is, the more frequently the strategy was used.**

In comparison with PSD principles, BCT taxonomy provides a more comprehensive pool of strategies for behavior change interventions. Even though, BCT taxonomy does not cover all the strategies we have found in related studies. Therefore, we add another five strategies to BCT taxonomy for our coding, which include social cooperation, social competition, social recognition, virtual reward, and game integration. The former three are derived from PSD principles. By game integration, we mean both exergames [63] and gamification [61].

---
[1] http://www.bct-taxonomy.com/interventions

## 2.3 The Behavioral Intervention Characteristics

The behavior change strategies are only about "what" but not "how" of the intervention. In 2014, Mohr and colleagues proposed the behavioral intervention technology (BIT) model to support the translation from behavior change intervention aims into an implementable treatment model [62]. Inspired by the BIT model, we include the concepts of intervention characteristics and intervention workflow in our coding, which can help us analyze how interventions are delivered. The characteristics include the medium, the visualization method (related to aesthetics), and the social support type in our coding. We further elaborate the details in the following section.

## 2.4 The Holistic Framework

In prior work, we proposed a holistic framework to guide the design and report of health behavior change interventions [88]. This framework integrates the three mentioned aspects in this section. By following this framework, we aim to provide the most comprehensive review of health behavior change in HCI. We emphasize comprehensiveness and consistency in reviewing health behavior change because health behavior change studies are always complex processes and affected by many aspects in field studies.

## 2.5 Related Work

In a highly related work in 2016, Orji and Moffatt [73] reviewed how persuasive technology was used for health and wellness in 85 related papers. They coded the reviewed studies from 11 perspectives: targeted (health) domain, technology, duration of evaluation, behavior theories, motivational strategies, evaluation method, targeted age group, number of participants, study country, targeted behavioral or psychological outcome, and findings/results. However, the coding of motivational strategies did not follow any existing taxonomy of persuasive technology (e.g., PSD principles) or behavior change techniques (e.g., BCT taxonomy). Thus the definitions of these strategies can be vague and imprecise for readers. We use a taxonomy integrating BCTs and PSD principles to code and analyze the adopted digital health strategies. Using the holistic framework [88] can help to achieve a more comprehensive review of the related studies.

Since existing systematic review on the effectiveness of digital health interventions have pointed out that there are not enough high-quality studies to draw powerful conclusion on effectiveness - e.g., eating behavior change [57] and sedentary behavior change at work [90] - we put our effort on revealing the patterns and trends of the existing empirical studies. We focus on multiple health behaviors instead of a specific one because we want to find out the patterns in different target behaviors.

## 3 Methods

As our initial aim is to find the research trends and patterns of health behavior change in the HCI community, we only used ACM digital library as our search repository, which covers most of HCI conference proceedings (e.g., CHI and UbiComp). For the searching, we considered the spelling versions of behavior/behaviour, similar expressions of behavior/behavioral change, persuasive technology, and the names of targeted behavior (e.g., physical activity and alcohol). We also excluded the papers focusing on sustainability, since they are out of the scope of this paper. The search was conducted on 23rd Aug. 2018, and the query syntax we used in ACM digital library is shown in Table 1.

**Table 1: The query syntax used in ACM digital library.**

*keywords.author.keyword:(+behavior +change -sustainability) OR keywords.author.keyword:(+behavioral +change -sustainability) OR keywords.author.keyword:(+behaviour +change -sustainability) OR keywords.author.keyword:(+behavioural +change -sustainability) OR keywords.author.keyword:(+persuasive +technology -sustainability) OR keywords.author.keyword:(+physical +activity) OR keywords.author.keyword:(diet dietary) OR keywords.author.keyword:(+sexual +health) OR keywords.author.keyword:(smoking) OR keywords.author.keyword:(sleeping) OR keywords.author.keyword:(sedentary sitting) OR keywords.author.keyword:(alcohol)*

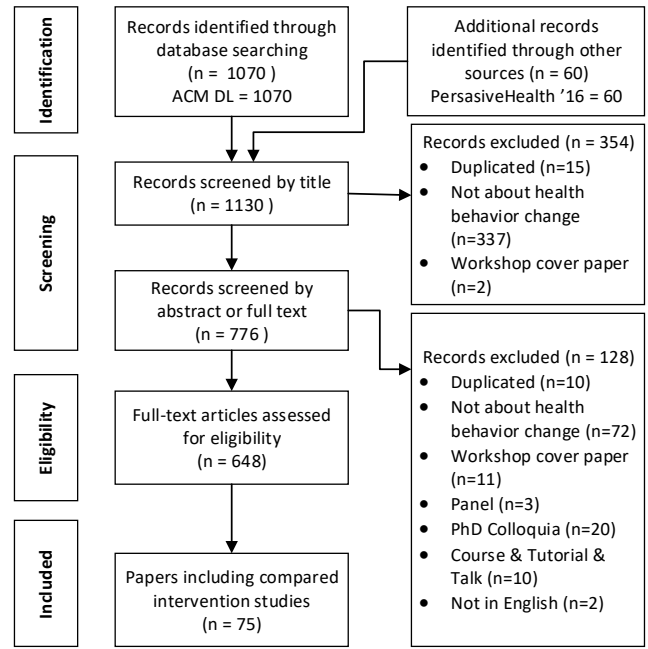

**Figure 3: The workflow of screening and selecting papers.**

The four-phase flow diagram of PRISMA [49] was used to illustrate the study selection process (see Figure 3). A total of 1070 papers were identified out of 530,358 records in ACM digital library. The first author screened the records from the paper title. 354 records were screened out because of duplication (N=15), not being relevant to health behavior change (N=337), or being the workshop introduction (N=2). Afterward, the first author reviewed the abstracts (or full paper if necessary) for the rest of the papers (N=776) and labeled the papers by the research method (see Table 2 for details) and target behavior. We further excluded the papers of duplication (N=10), not about health behavior change (N=72), workshop introductions (N=2), panel introductions (N=3), Ph.D. colloquia (N=20), courses & tutorials & talk introductions (N=4), and not in English (N=2). Finally, we obtained 648 papers falling into the listed types in Table 2. The paper list can be found in the supplementary material.

**Table 2: The paper types used in our coding.**

| Type | Explanation |
|---|---|
| Compared intervention study | It includes at least one user intervention session with at least two compared conditions. |
| Exploration study | It includes at least one user intervention session for behavioral factors exploration without compared conditions |
| Test study | It is a feasibility test or pilot study without enough measures of users' behavior or outcome. |
| Design | It is about designing systems or methods to support health behavior change without any user intervention. |
| Interview | It includes only user interviews without any user intervention. |
| Survey | It includes only surveys based on questionnaires without any intervention to users. |
| Data mining | It is about systems/algorithms to detect, recognize, classify, or predict human behavior or behavioral factors for health behavior change. |
| Review | It overviews or reviews previous work. |
| Framework & Theory | It proposes frameworks or theories for the research on health behavior change. |
| Viewpoint | It provides viewpoints, guidelines, or implications for the research on health behavior change. |
| Concept | It includes only concepts of systems or methods for health behavior change. |

From the 648 papers in the phase of eligibility, we selected 72 papers that include compared intervention studies (73 studies in total). Afterward, two of the authors coded these full papers separately, and the differences were resolved by discussion. The coding schema is shown in Table 3.

## 4 Results and Findings

In this section, we report the results and findings of the systematic review based on the methods introduced in the previews section. We firstly show the trends of the paper amount in the perspectives of the target behavior and the paper type (see Table 2) over the research history of health behavior change in HCI. Then we select "compared intervention studies" (N=75) and analyze the research patterns in the views of measurements, user experience evaluation, the target behavior, the target user group, the application of behavioral theories, the use of behavior change strategies, and intervention characteristics.

**Table 3: The coding items and explanations.**

| No. | Item | Explanation |
|---|---|---|
| 1 | Target behavior | Physical activity, diet, etc. |
| 2 | Target user group | Adults, children, etc. |
| 3 | Behavioral theory | TTM, goal-setting theory, etc. |
| 4 | Behavioral theory use type | Informing design; guide evaluation strategies; selecting target users. |
| 5 | Behavior change strategy | BCT Taxonomy (v1) + (cooperation, competition, Recognition, virtual reward, and game integration). |
| 6 | Measurement | User experience (quantitative); user experience (qualitative); target behavior; user interaction (i.e., use frequency, use duration); behavioral factors (e.g., constructs from behavioral theories). |
| 7 | User experience Instrument | SUS [12], AttrakDiff[2], etc. (Coding only when user experience is quantitatively measured.) |
| 8 | Intervention workflow | Time-based; task-based; event-based. (Coding only when prompts/cues are used as a behavior change strategy.) |
| 9 | Intervention Characteristic | See Table 4 for details. |

**Table 4: The characteristics used in our coding and the explanations.**

| No. | Characteristic | Explanation |
|---|---|---|
| 1 | Medium | The device for intervention delivery, e.g., PC and smartphone. |
| 2 | Visualization | Information visualization in software interfaces, e.g., progress bar and leaderboard. |
| 3 | Social support | The social support type that the intervention can aid, e.g., social comparison and social recognition. See Table 5 for details. (Coding only when the intervention system provides social support function) |

### 4.1 Trends of Target Behaviors and Paper Types

We analyze the trends based on the papers after the title and abstract screening (N=648). Of the 48 target behaviors we found, five behaviors (i.e., physical activity, sleep, diet, smoking, and sedentary behavior) account for 73% in all the papers. As shown in Figure 4, physical activity remains the most popular target behavior over time and the corresponding papers keep growing in the last six years. The number of the papers targeting sedentary behavior also peaked in 2017. The paper amounts for sleep, diet, and others decreased in 2017 after 2-4 years increase. From the perspective of the target behavior, the decrease of papers about sleep, dietary behavior, and other behaviors except the ones listed in Figure 4 caused the drop in the overall paper count in 2017.

---
[2] http://attrakdiff.de

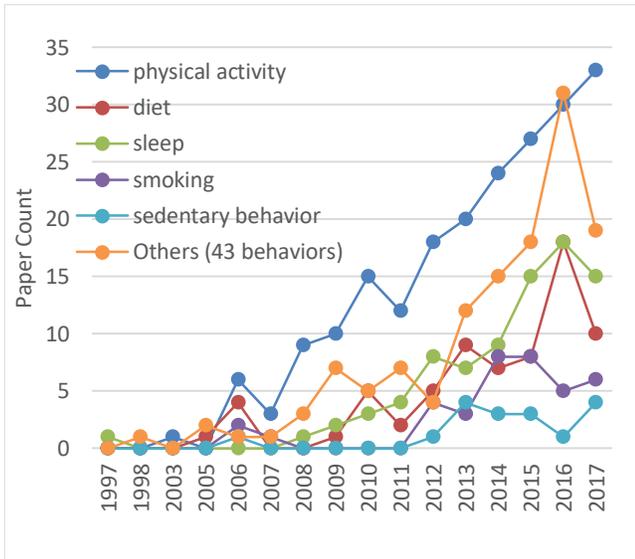

**Figure 4: The target behaviors in the reviewed papers over time (N=648).**

Figure 5 illustrates the change of the paper amount regarding the paper type over time. Most (55%) of the papers contain designing new intervention systems or methods for health behavior change. However, only about 25% of the developed systems or methods were evaluated by the intervention study with compared conditions. The "data mining" papers saw a drop in 2017, while the "survey" papers and the "interview" papers have been rising in the last three years. The drop in "data mining" and "design" papers mainly contributed to the overall decrease in 2017.

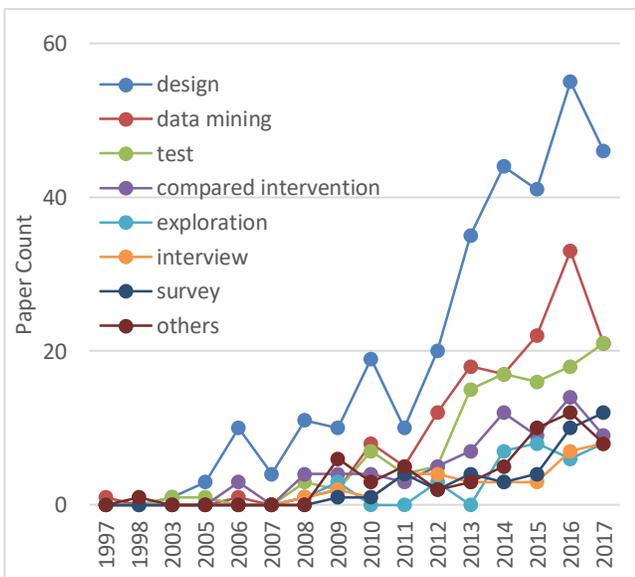

**Figure 5: The paper category in the reviewed papers over time (N=648).**

## 4.2 Patterns in the Selected Intervention Studies

### 4.2.1 Measurement & User Experience Evaluation

Differing from the target behavior as shown in Figure 4, the user interaction means the objective measure of how the users use the intervention system (e.g., the use frequency). In comparison with the user experience, the behavioral factors refer to the constructs (e.g., self-efficacy) influencing the behavior change process. The target behavior of users was measured in most of the studies (93%), as shown in Figure 6. More than half of the studies (59%) evaluated user experience quantitatively or qualitatively. The user interaction with the intervention system was measured in 34% of the studies. Only 20% of the studies accessed users' behavioral factors, which is related to the usage of behavioral theories.

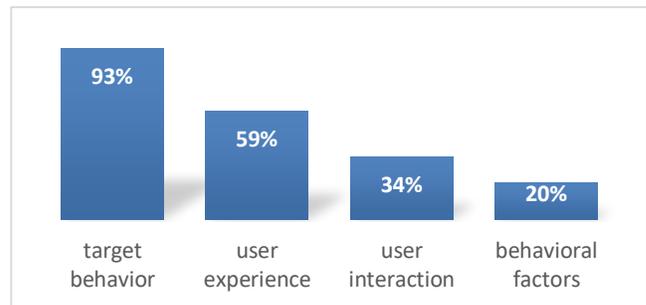

**Figure 6: The percentages for the used measurements in the reviewed studies.**

Although about 59% of the studies accessed the user experience, only 32% (24/75) of them evaluated the user experience with quantitative measurements. The system usability scale (SUS) was used in four studies [25,26,55,95], while the NASA-TLX [45] and the AttrakDiff [21] were used only once. The studies with game integration were more likely to measure users' perceived enjoyment (e.g., [8,37,56]). One study [94] was conducted within a clinical trial, which used the Patient Reaction Assessment (PRA) questionnaire to measure users' experience of the intervention. We did not find any specific scale to evaluate the user experience of interventions for health behavior change.

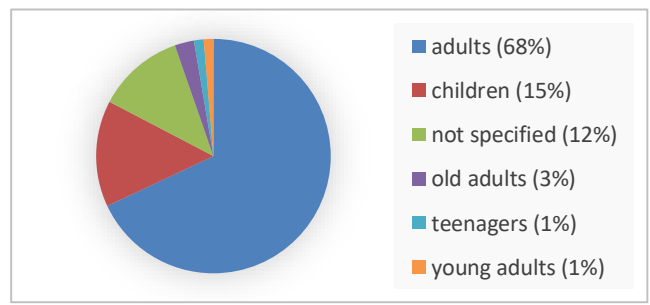

**Figure 7: The distribution of the target user group.**

### 4.2.2 Target User Group & Behavior

As shown in Figure 7, the target user group in most studies was the adult (68%), while almost half of these studies used college students and staff as the participants. Children, as the target user group, accounted for 15% in all studies. There is only one study targeting teenagers, while one study focused on young adults. From other aspects of the user group: five studies aimed at patients, three studies focused on the female, and one study only considered athletes. The reviewed studies are very unbalanced regarding the target user group.

*4.2.3 Behavioral Theories*

Among the 75 selected intervention studies, 32 studies (43%) explicitly described the use of behavioral theories. The transtheoretical model (TTM) was the most frequently used theory, which was adopted in eight papers. This result is in line with another systematic review by Orji and Moffatt [73]. The other behavioral theories adopted in the reviewed studies are listed in Figure 8.

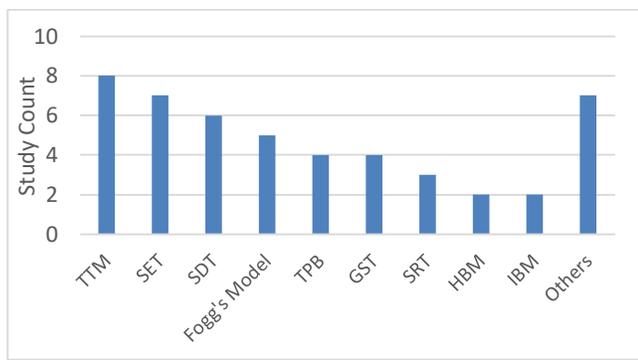

**Figure 8: The distribution of the used behavioral theories. SRT refers to the self-regulation theory; HBM refers to the health belief model; IBM refers to the integrated behavioral model. The other acronyms can be found in the following content.**

Figure 9 illustrates how behavioral theories were used in the reviewed studies. The TTM was mainly used to select target users [18,19,35,40,45,50], as illustrated in [38]. In the case of using the TTM to inform the intervention design, different strategies were delivered according to the user' stage of change [36,39]. The self-efficacy theory (SET) [4], the theory of planned behavior (TPB), the self-determination theory (SDT) [22], the Fogg's behavior model [30], and the goal-setting theory (GST) [52] largely contributed to informing the intervention design.

Regarding the studies that did not utilize behavioral theories, we found that 29% (12 studies) focused on exergame and gamification (e.g., [82]), while 21% (9 studies) targeted children or teenagers (e.g., [86]). Behavioral theories might be useless in the case of the short game session (e.g., exergame). However, users' adoption and engagement with health orientated game design could also be explained by behavioral theories. The work from Macvean and Robertson [55] indicated that children's motivation of playing exergame would decrease over time and self-efficacy theory can predict and interpret the longitudinal physical activity patterns of children's behavior change as well.

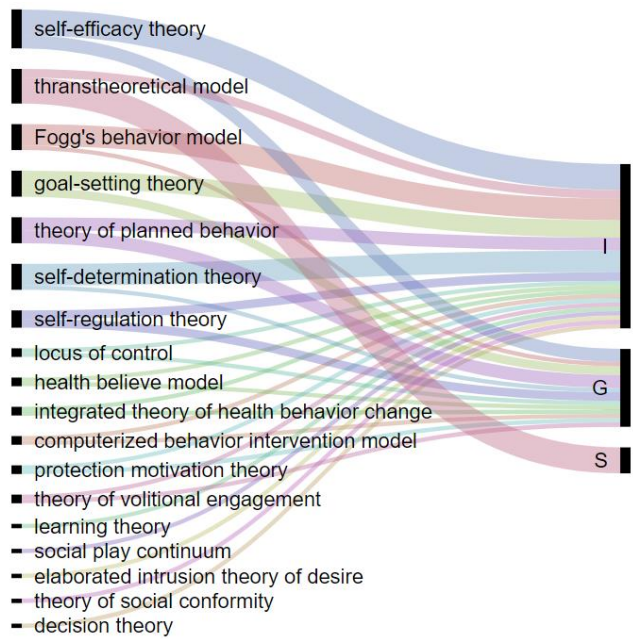

**Figure 9: Behavioral theories and the ways that they were used in the reviewed studies. I – informing design, G – guide evaluation strategies, S – selecting target users. Note that, in this alluvial graph, the relative sizes of the bars for each behavioral theory do not exactly represent their use frequency. In one study, a behavioral theory can be used for both informing design and guiding evaluation strategies.**

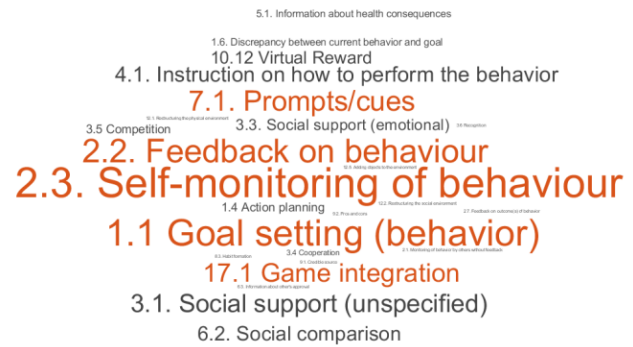

**Figure 10: The word cloud of behavior change strategies found in our reviewed intervention studies (N=75). The bigger the font is, the more frequently the strategy was used.**

*4.2.4 Behavior Change Strategies*

Among the reviewed studies, the most used behavior change strategies are self-monitoring of behavior, goal-setting (behavior), feedback on behavior, prompts/cues, and game integration (see Figure 10). In section 1.2, we have shown the cloud map of behavior change strategies coded from the papers listed on the website of BCT taxonomy. Those papers are mainly from journals of public health, behavioral science, and healthcare (e.g., BMC public health[3] and JMIR[4] ). In comparison with our reviewed papers, the researchers of those papers are more likely to use goal-setting (behavior), action planning, problem-solving, instruction on how to perform a behavior, and information about health consequences. This indicates the different use patterns of behavior change strategies between the HCI community and the community of public health, behavioral science, and healthcare.

We have shown the distribution of the target behavior in the papers of all the selected types (N=648) in Figure 4, where the papers targeting physical activity are much more than other types. Among the 75 full coded studies, the target behaviors are also unbalanced in quantity, and physical activity is still the most addressed behavior. Figure 11 illustrates the interaction of the top-4 target behaviors and behavior change strategies (in-group) in the reviewed studies. The alluvial graph indicates that a variety of behavior change strategies was used for all the target behaviors. One interesting finding is that almost all of the studies using game integration were designed for physical activity.

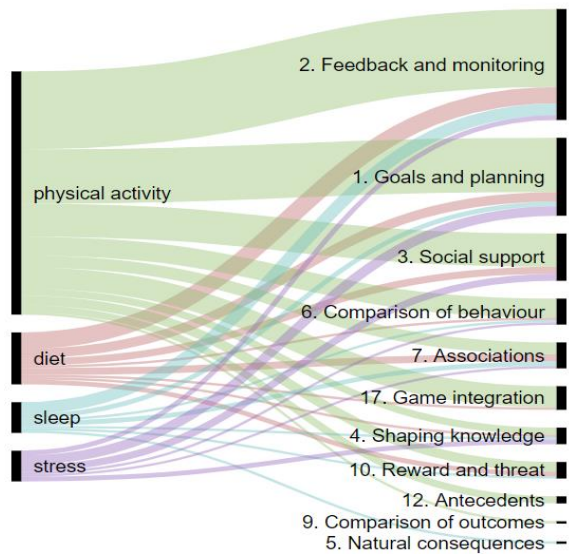

**Figure 11: The top-4 target behaviors and the corresponding behavior change strategies groups in the reviewed studies. Note that, in this alluvial graph, the relative sizes of the bars for each target behavior do not exactly represent their frequency. Given the target behavior in one study, several strategies might be used.**

*4.2.5 Intervention Workflow*

We found 16 out of 75 studies involving reminders (i.e., prompts/cue), including 9 event-based reminders [9,10,13,21,26,48,83,84], and 7 time-based reminders [7,32,44,66,75,85,94]. We did not find any task-based intervention workflow, according to the definition in the BIT model (i.e., the release of intervention elements are based on the user's completion of prescribed intervention tasks [62]). Among the studies where the intervention system did not provide any scheduled reminders or prompts, we found a group of studies using always-on glanceable cues [5,19,35,78] to nudge users. For example, Gouveia and colleagues [35] designed watch faces of the smartwatch to provide real-time feedback about the user's physical activity.

### 4.3 Characteristics of the Selected Intervention Studies

*4.3.1 Media*

The media determine the information channel of the intervention. The mobile phone (including the smartphone and the functional phone) was used in most of the studies (44/75). The mobile phone, especially the smartphone, has become indispensable in our daily life. Therefore, the high adoption rate of the mobile phone is not surprising. The rest of the adopted media in the studies are listed in Figure 12. The web means that the study did not restrict users to use a mobile device or a PC. Except for the common devices (e.g., PC, the mobile phone, the fitness tracker), some new devices were created to solve specific problems. For example, the wearables for monitoring sitting poster [26] and augmented slider for supporting children's learning process [56].

*4.3.2 Visualization*

The visualization means how the intervention is presented to the users via the software interface. As shown in Figure 13, the plain text (e.g., SMS and notification), the progress bar, and the gamification interface were the most popular visualization methods. The others include the virtual agent, the timeline, the leaderboard, the reward sheet, the icon, the cartoon figure, the Emoji, and so on.

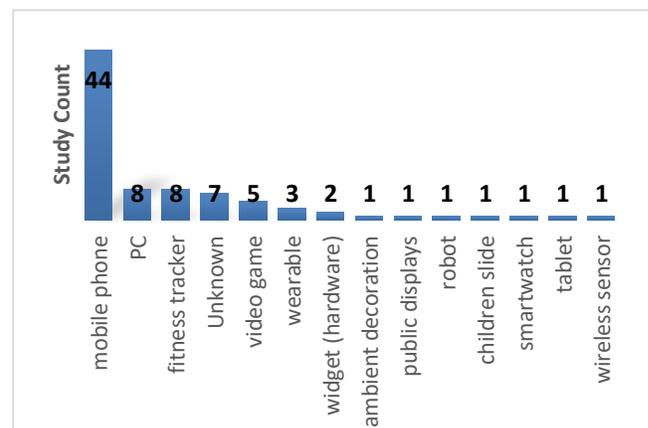

**Figure 12: The media used in the reviewed studies. The unknown refers to the studies that did not explicitly mention the media. The wearable means the ones users can wear on clothes or shoes, rather than fitness trackers and smartwatches.**

---
[3] https://bmcpublichealth.biomedcentral.com/

[4] http://www.jmir.org/

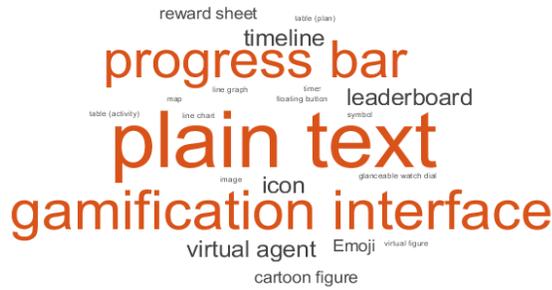

**Figure 13: The word map of the visualization methods used in the reviewed studies. The bigger the font it, the more frequently the visualization method was used.**

*4.3.3 Social Support*

We found 21 studies that provided the function of social support. The types of social support appeared in these studies are listed in Table 5. Some of the types are included in the BCT taxonomy [59] (e.g., social comparison and social incentive) and PSD principles [70] (e.g., social cooperation and social competition). However, our social support types and the explanations might be different from the definitions in the BCT taxonomy and PSD principles. Instead, we derived these social support types by analyzing the intervention descriptions in the reviewed studies.

**Table 5: The social support types.**

| No. | Social Support Type | Explanation |
|---|---|---|
| 1 | Social commitment | It allows a user to make commitments within the intervention platform [66]. |
| 2 | Social sharing | It allows users to see each other's status, without aiming to compare with each other within the intervention platform [25,42]. |
| 3 | Social comparison | It allows users to compare with each other within the intervention platform [16,18,28]. |
| 4 | Social competition | It allows users to compete with each other within the intervention platform [31,37,68,95]. |
| 5 | Social communication | It allows users to communicate with each other within the intervention platform [11,18]. |
| 6 | Social incentive | It allows users to encourage each other within the intervention platform [16,25,41,83]. |
| 7 | Social interaction | It allows users to interact with each other with the intervention tool face to face [6,53]. |
| 8 | Social monitoring | It allows other users to monitor a user's behavior, but not vice versa [75,83,93]. |
| 9 | Social recognition | It allows users to recognize their peers in public [43,78]. |

## 5  Discussion

We have reported our findings of the research trends and patterns of the research on health behavior change in the HCI community. These findings indicate several shortcomings and problems to be addressed in this research field:

1. The target behaviors mainly fell into physical activity, while some critical behaviors (e.g., sedentary behavior [91] and stress management [81]) were much less addressed.
2. The target users or study participants were mostly adults who were mostly college students and staff. We believe that more attention on the elderly is required in the aging society.
3. There is no standardized approaches or instruments for evaluating the user experience of intervention systems for health behavior change.
4. There is no standard to report the intervention study for health behavior change. The method we used to review the related papers provide an approach to reporting the relevant study. The study aspects that we suggest to report are shown in Table 3 and Table 4.

Following, we select six topics inspired by the reviewed studies to provide suggestions and opportunities for future studies. The first three topics are about users: considering the user's behavior priority, categorizing target users from different perspectives, and leveraging users' power of creativity and engagement. The other topics include longitudinal studies with game integration, cautions for socialization, and the applications of AI in health behavior change.

### 5.1  Users' Behavior Priority

Although behavioral theories could be beneficial for the research on health behavior change in HCI, they are not without limitations. One of the limitations is that behavioral theories can explain only 20-30% of the total variance in a given health behavior [38]. From the reviewed papers, we noticed one factor that could collaboratively explain health behavior change. In the study by Rodgers and colleagues [77], they found that college students consciously prioritize academic success over a healthy sleeping pattern. This finding indicates the fact that an individual's daily life is filled with various tasks and behaviors (e.g., academic success and healthy sleeping pattern), instead of only the target behavior of a given intervention study. People can fail to adhere to an action plan just because they need to do other actions with higher priority in their limited time. Systems that can support users to schedule their daily activity and fit the target behavior into their routine could be a solution to the difficulty caused by priority.

Therefore, we suggest that future intervention studies should consider the behavior priority of participants. For example, the efficacy of sedentary behavior interventions could relate to users' work priority. Intervention designers should check if there are critical tasks or dues hindering users' enactment of their plans to reduce their sedentary behavior.

## 5.2 Categorizing Target Users

An intervention may be valid only for a specific group of audience, and the lack of specification of users could hide the effectiveness in study results. For example, Lacroix and colleagues [46] found that positive linear relationship between goal difficulty and users' performance only existed for inactive people, but not for active people. Therefore, categorizing the target users in meaningful ways could lead to a better understanding of the intervention efficacy. In Figure 9, we have shown that the transtheoretical model was often used to group participants into different stages of change and select the target users. Besides the stage of change, researchers have found other methods and perspectives to categorize target users. E.g., Wiafe and colleagues [92] proposed a model to classify users based on their current behavior, attitude, and levels of cognitive dissonance.

Users' personality is also a potential way to categorize target users, e.g., the well-known Big Five personality traits [87]. As many health intervention studies have used gamification, the research on the personality of users (players) in games has drawn more attention [67]. Orji et al. [74] examined how different personalities respond to various persuasive strategies that are used in persuasive health games and gamified systems. Another study [72] showed that tailoring the game design to players' personality type improved the effectiveness of the games in promoting positive attitudes, intention to change behavior, and self-efficacy.

Our second suggestion is that future intervention could categorize target users not only based on the stage of change but also other factors (e.g., personality).

## 5.3 Leveraging Users' Power

Researchers have started to explore and leverage the users' creativity in health behavior change. Lee and colleagues [47] deployed a self-experiment design to support behavior change for improving participants' sleep quality. In another study [75], a participatory design session was used after an intervention session for medication management among the elderly. Both of the studies proved the efficacy of user participation in the intervention design process.

In the study of Birk and Mandryk [10], a group of users was asked to customize their avatars to interact in a breathing exercise program. Compared to the control group with randomly assigned avatars, the customization group saw significantly less attrition and more sustained engagement through the 3-week study. In this study, customizing an avatar with its appearance and attributes required a minimum of 4-minute work of a participant. The effect of users' participation can be explained not only from the perspective of customization but also from the view of IKEA effect [69,89]. The involvement of users' effort in a product can increase their evaluation of the product.

Based on the evidence, we suggest that future intervention designers should take advantage of users' participation and further explore the effect and user experience of participation.

## 5.4 Longitudinal Study with Game Integration

By game integration, we mean both exergames (i.e., interactive games that require players to invest significant physical effort as part of the gameplay [63]) and gamification (i.e., implementing the most common and enjoyable mechanics of video games in non-video game contexts) [61]. We extracted 15 studies with game integration and found that the study duration tended to be short, as shown in Figure 14. Macvean and Robertson [55] studied children's physical activity patterns when using an exercise game on smartphones over seven weeks, which is the longest study on gamification among the 15 studies. Seven studies only reported their evaluation for one game session, which we counted as one day in Figure 14.

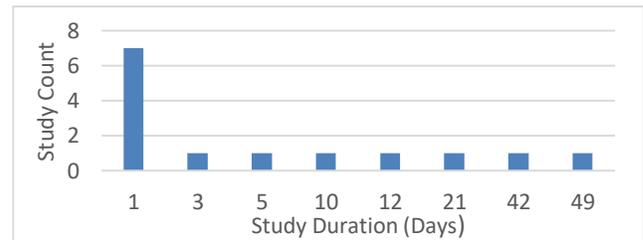

Figure 14: The study duration of the studies with game integration.

The distribution of the reviewed study durations is shown in Figure 15, where we can see the number of studies with the duration less than one day is 13. More than half of the short-term studies are about game integration. Therefore, more longitudinal studies in gamification are required, because the goal of health behavior change is to help users maintain a healthy lifestyle in the long term.

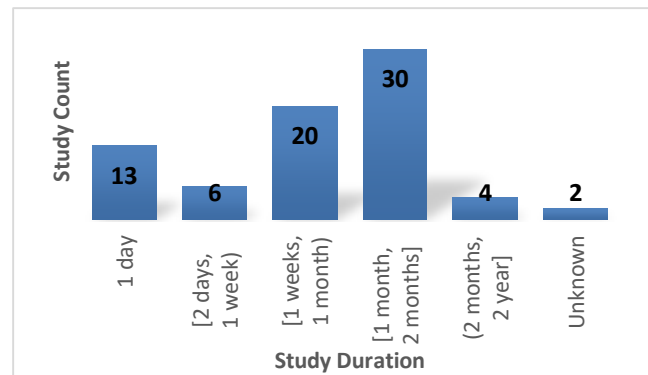

Figure 15: The distribution of the reviewed study durations.

## 5.5 Cautions for Socialization

Since every individual is part of the social network and mobile technologies keep changing our communication in the social network, it is vital to investigate how socialization can benefit health behavior change. Socialization for health behavior change means involving the support from private (e.g., families and friends) or public social networks in health interventions. We have listed the social support types in the review studies in Table 5.

Katule et al. targeted parent-child pairs to improve the parent's physical activity via the child [41]. Chen and colleagues [15] found that collaborating with a buddy (dyads) to compete in a community can be effective to improve the daily steps of obese and diabetic patients. While studies have shown that social incentives have the potential to motivate people for health behavior change, some research showed cautions when deploying socialization. E.g., social interactions could be demotivating between dyads who did not know each other well [15]. The work of Munson et al. [66] showed that the prospect of public accountability might suppress the making of commitments which decreases social members' motivation of obese adults.

More research on the requirements of socialization for different user groups is needed. For instance, how to apply social support to improve physical activity and dietary behavior of the elderly living along?

### 5.6 Embracing AI

Although recent applications of deep learning have boomed in many fields, its use for health behavior change is still in infancy. By AI in health behavior change, we refer to the system that adopts a social role in communicating with users for health behavior change. This definition emphasizes the interaction between the system and users. It excludes the system only providing functional support, e.g., food and ingredient recognition [14]. We found three systems following our definition of AI in health behavior change. Kanaoka and Mutlu used a humanoid robot to motivate users for physical activity in two interaction conditions [40]. Over a two-week study, although no significant difference was found in the users' physical activity level, their intrinsic motivation was significantly improved. Interestingly, users' willingness and perceived friendliness of the robot are both higher in the monologue condition (less interaction) than in the motivational interviewing condition (more interaction). This result might be due to the lack of fluency and error in speech recognition [40]. Another system by Lisetti and colleagues developed a virtual agent to deliver interventions on excessive alcohol consumption [51]. Their virtual agent can recognize users' expressions and generate corresponding expressions to show empathic feedbacks. With empathic feedbacks, the virtual agent improved users' attitude to the technology, intention to use, perceived enjoyment, and so on. Unlike the mentioned two systems with anthropomorphism, Kamphorst and colleagues developed an autonomous e-coaching system to deliver intervention messages to promote more stairs taking according to the problematic constructs for behavior change of users in real-time [39]. A month-long evaluation study showed that the intelligent e-coaching system could better support health behavior change.

The initial results of applying AI technology implicate the venues of research on health behavior change in HCI: natural language based intervention [40,51], emotion enabled intervention [27,51], and computational intervention [39].

### 6 Limitations

Several limitations emerged during the systematic review. The search was conducted only in the ACM digital library, so the reviewed papers did not include all the related work in the HCI community. This might lead to bias in our results. We only searched the authors' keywords to extract papers, which may also lead to missing some related papers. No paper explicitly reported the intervention strategies based on the taxonomy of behavior change techniques (BCTs). The coding is based on the authors' understanding of the material provided on the website of BCTs taxonomy[5], which might introduce bias to our analysis.

### 7 Conclusion

Through a systematic review of the research on health behavior change in the HCI community, this paper shows the research trends in the perspectives of the target behavior and paper types over the research history (N=648). Based on the selected intervention studies (N=75), it also analyzes the research patterns in the views of measurements, user experience evaluation, the target behavior, the target user group, the use of behavioral theories, the use of behavior change strategies, and intervention characteristics. The results show that physical activity was the most targeted behavior over time, and the related research keeps growing in recent years. Other behaviors (e.g., sleep, dietary behavior, smoking, and sedentary behavior) increasingly draw more attention with slight fluctuations. Studies using interviews or surveys continue to increase, while research on data mining and designing new intervention systems or methods dropped in 2017. Among the 75 intervention studies with compared conditions, only 32% of the studies quantitatively evaluated the user experience. The SUS, the NASA-TLX, and the AttrakDiff were used in the reviewed studies, while no standardized method to assess the user experience of intervention studies for health behavior change was found. Most of the target users in these studies were adults. There were 32 out of 75 studies explicitly reporting the use of behavioral theory, and the most used one is the transtheoretical model. A variety of behavior change strategies were used in the reviewed studies, while the most frequently used ones include self-monitoring of behavior, goal-setting (behavior), feedback on behavior, prompts/cues, and game integration. Almost all the studies with game integration were designed for physical activity. The mobile phone was the most popular medium to deliver interventions. The plain text, the progress bar, and the gamification interface were the top-3 visualization methods to provide information to the user. Regarding social support, we found nine use cases among the reviewed studies. Based on these findings, we discuss the shortcomings and problems to be addressed: unbalanced target behaviors, unbalanced target user groups, the lack of standardized evaluation methods for user experience, and the lack of standards to report intervention studies for health behavior change in HCI.

Finally, we provide suggestions and opportunities for the future research in the field of health behavior change in HCI. We suggest considering users' behavior priority and the ways to categorize target users when recruiting the study participants. We also point out the lack of longitudinal studies for game-integrated systems and the cautions for socialization-engaged systems. Also, we show how AI technologies have been used for health behavior change and implicate the research venues of AI in this field.

---

[5] http://www.bct-taxonomy.com/dashboard

Responding to the trend of decrease in the related paper amount from the HCI community, we believe it is a temporary phenomenon. According to the findings and analysis in this paper, there are many unexplored research questions and opportunities in health behavior change for HCI researchers.

# REFERENCES

1. Charles Abraham and Susan Michie. 2008. A taxonomy of behavior change techniques used in interventions. *Health Psychology* 27, 3: 379–387. https://doi.org/10.1037/0278-6133.27.3.379
2. Icek Ajzen. 1991. The theory of planned behavior. *Orgnizational Behavior and Human Decision Processes* 50: 179–211. https://doi.org/10.1016/0749-5978(91)90020-T
3. Albert Bandura. 1977. Toward a unifying theory of behavioral change. *Psychological Review 84*, 191–215. https://doi.org/10.1037/0033-295X.84.2.191
4. Albert Bandura and Nancy E. Adams. 1977. Analysis of self-efficacy theory of behavioral change. *Cognitive Therapy and Research* 1, 4: 287–310. https://doi.org/10.1007/BF01663995
5. Jared S Bauer, Sunny Consolvo, Benjamin Greenstein, Jonathan Schooler, Eric Wu, Nathaniel F Watson, and Julie Kientz. 2012. ShutEye: Encouraging Awareness of Healthy Sleep Recommendations with a Mobile, Peripheral Display. In *Proceedings of the SIGCHI Conference on Human Factors in Computing Systems* (CHI '12), 1401–1410. https://doi.org/10.1145/2207676.2208600
6. Tilde Bekker, Janienke Sturm, and Berry Eggen. 2010. Designing Playful Interactions for Social Interaction and Physical Play. *Personal Ubiquitous Comput.* 14, 5: 385–396. https://doi.org/10.1007/s00779-009-0264-1
7. Frank Bentley and Konrad Tollmar. 2013. The power of mobile notifications to increase wellbeing logging behavior. In *Proceedings of the SIGCHI Conference on Human Factors in Computing Systems - CHI '13* (CHI '13), 1095. https://doi.org/10.1145/2470654.2466140
8. Shlomo Berkovsky, Mac Coombe, Jill Freyne, Dipak Bhandari, and Nilufar Baghaei. 2010. Physical Activity Motivating Games: Virtual Rewards for Real Activity. In *Proceedings of the SIGCHI Conference on Human Factors in Computing Systems* (CHI '10), 243–252. https://doi.org/10.1145/1753326.1753362
9. Dave Berque, Jimmy Burgess, Alexander Billingsley, ShanKara Johnson, Terri L Bonebright, and Brad Wethington. 2011. Design and Evaluation of Persuasive Technology to Encourage Healthier Typing Behaviors. In *Proceedings of the 6th International Conference on Persuasive Technology: Persuasive Technology and Design: Enhancing Sustainability and Health* (PERSUASIVE '11), 9:1--9:10. https://doi.org/10.1145/2467803.2467812
10. Max V. Birk and Regan L. Mandryk. 2018. Combating Attrition in Digital Self-Improvement Programs using Avatar Customization. In *CHI '18: Proceedings of the SIGCHI Conference on Human Factors in Computing Systems: Proceedings of the SIGCHI Conference on Human Factors in Computing Systems* (CHI '18), 1–15. https://doi.org/10.1145/3173574.3174234
11. Ludovico Boratto, Salvatore Carta, Fabrizio Mulas, and Paolo Pilloni. 2017. An e-Coaching Ecosystem: Design and Effectiveness Analysis of the Engagement of Remote Coaching on Athletes. *Personal Ubiquitous Comput.* 21, 4: 689–704. https://doi.org/10.1007/s00779-017-1026-0
12. John Brooke. 2013. SUS: a retrospective. *Journal of Usability Studies* 8, 2: 29–40. Retrieved September 19, 2018 from https://dl.acm.org/citation.cfm?id=2817913
13. Ana Caraban, Evangelos Karapanos, Pedro Campos, and Daniel Gonçalves. 2018. Exploring the Feasibility of Subliminal Priming on Web Platforms. In *Proceedings of the 36th European Conference on Cognitive Ergonomics* (ECCE'18), 8:1--8:11. https://doi.org/10.1145/3232078.3232095
14. Jingjing Chen and Chong-wah Ngo. 2016. Deep-based Ingredient Recognition for Cooking Recipe Retrieval. In *Proceedings of the 2016 ACM on Multimedia Conference - MM '16*, 32–41. https://doi.org/10.1145/2964284.2964315
15. Yu Chen, Mirana E. Randriambelonoro, Antoine Geissbuhler, and Pearl Pu. 2016. Social Incentives in Pervasive Fitness Apps for Obese and Diabetic patients. In *Proceedings of the 19th ACM Conference on Computer Supported Cooperative Work and Social Computing Companion - CSCW '16 Companion* (CSCW '16 Companion), 245–248. https://doi.org/10.1145/2818052.2869093
16. Meng-chieh Chiu, Shih-ping Chang, Yu-chen Chang, Hao-hua Chu, Cheryl Chia-hui Chen, Fei-hsiu Hsiao, and Ju-chun Ko. 2009. Playful Bottle : a Mobile Social Persuasion System to Motivate Healthy Water Intake. *Proceedings of the 11th international conference on Ubiquitous computing*: 184–194. https://doi.org/10.1145/1620545.1620574
17. David E. Conroy, Chih-Hsiang Yang, and Jaclyn P. Maher. 2014. Behavior Change Techniques in Top-Ranked Mobile Apps for Physical Activity. *American Journal of Preventive Medicine* 46, 6: 649–652. https://doi.org/10.1016/j.amepre.2014.01.010
18. Sunny Consolvo, Katherine Everitt, Ian Smith, and James A Landay. 2006. Design Requirements for Technologies That Encourage Physical Activity. In *Proceedings of the SIGCHI Conference on Human Factors in Computing Systems* (CHI '06), 457–466. https://doi.org/10.1145/1124772.1124840
19. Sunny Consolvo, Predrag Klasnja, David W McDonald, Daniel Avrahami, Jon Froehlich, Louis LeGrand, Ryan Libby, Keith Mosher, and James A Landay. 2008. Flowers or a Robot Army?: Encouraging Awareness & Activity with Personal, Mobile Displays. In *Proceedings of the 10th International Conference on Ubiquitous Computing* (UbiComp '08), 54–63. https://doi.org/10.1145/1409635.1409644
20. David Crane, Claire Garnett, James Brown, Robert West, and Susan Michie. 2015. Behavior Change Techniques in Popular Alcohol Reduction Apps. *Journal of Medical Internet Research* 17, 5: e118. https://doi.org/10.2196/jmir.4060
21. Saskia Dantzig, Gijs Geleijnse, and Aart Tijmen Halteren. 2013. Toward a Persuasive Mobile Application to Reduce Sedentary Behavior. *Personal Ubiquitous Comput.* 17, 6: 1237–1246. https://doi.org/10.1007/s00779-012-0588-0
22. Edward L. Deci and Richard M. Ryan. 2000. The "What" and "Why" of Goal Pursuits: Human Needs and the Self-Determination of Behavior. *Psychological Inquiry* 11, 4: 227–268. https://doi.org/10.1207/S15327965PLI1104_01
23. Artur Direito, Leila Pfaeffli Dale, Emma Shields, Rosie Dobson, Robyn Whittaker, and Ralph Maddison. 2014. Do physical activity and dietary smartphone applications incorporate evidence-based behaviour change techniques? *BMC Public Health* 14, 1: 646. https://doi.org/10.1186/1471-2458-14-646
24. Stephan U. Dombrowski, Falko F. Sniehotta, Alison Avenell, Marie Johnston, Graeme MacLennan, and Vera Araújo-Soares. 2012. Identifying active ingredients in complex behavioural interventions for obese adults with obesity-related co-morbidities or additional risk factors for co-morbidities: a systematic review. *Health Psychology Review* 6, 1: 7–32. https://doi.org/10.1080/17437199.2010.513298
25. Honglu Du, G. Michael Youngblood, and Peter Pirolli. 2014. Efficacy of a Smartphone System to Support Groups in Behavior Change Programs. In *Proceedings of the Wireless Health 2014 on National Institutes of Health - WH '14* (WH '14), 1–8. https://doi.org/10.1145/2668883.2668887
26. Jiachun Du, Qi Wang, Liesbet de Baets, and Panos Markopoulos. 2017. Supporting shoulder pain prevention and treatment with wearable technology. In *Proceedings of the 11th EAI International Conference on Pervasive Computing Technologies for Healthcare - PervasiveHealth '17* (PervasiveHealth '17), 235–243. https://doi.org/10.1145/3154862.3154886
27. Ahmed Fadhil, Gianluca Schiavo, Yunlong Wang, and Bereket A. Yilma. 2018. The Effect of Emojis when interacting with Conversational Interface Assisted Health Coaching System. In *Proceedings of the 12th EAI International Conference on Pervasive Computing Technologies for Healthcare - PervasiveHealth '18*, 378–383. https://doi.org/10.1145/3240925.3240965
28. Andre T S Fialho, Herjan van den Heuvel, Qonita Shahab, Qing Liu, Li Li, Privender Saini, Joyca Lacroix, and Panos Markopoulos. 2009. ActiveShare: Sharing Challenges to Increase Physical Activities. In *CHI '09 Extended Abstracts on Human Factors in Computing Systems* (CHI EA '09), 4159–4164. https://doi.org/10.1145/1520340.1520633
29. B. J. Fogg. 2003. *Persuasive technology : using computers to change what we think and do*. Morgan Kaufmann Publishers.
30. Bj Fogg. 2009. A behavior model for persuasive design. *Proceedings of the 4th International Conference on Persuasive Technology - Persuasive '09*: 1. https://doi.org/10.1145/1541948.1541999
31. Derek Foster, Conor Linehan, Ben Kirman, Shaun Lawson, and Gary James. 2010. Motivating Physical Activity at Work: Using Persuasive

Social Media for Competitive Step Counting. In *Proceedings of the 14th International Academic MindTrek Conference: Envisioning Future Media Environments* (MindTrek '10), 111–116. https://doi.org/10.1145/1930488.1930510
32. Jill Freyne, Emily Brindal, Gilly Hendrie, Shlomo Berkovsky, and Mac Coombe. 2012. Mobile Applications to Support Dietary Change: Highlighting the Importance of Evaluation Context. In *CHI '12 Extended Abstracts on Human Factors in Computing Systems* (CHI EA '12), 1781–1786. https://doi.org/10.1145/2212776.2223709
33. Benjamin Gardner, Lee Smith, Fabiana Lorencatto, Mark Hamer, and Stuart JH Biddle. 2016. How to reduce sitting time? A review of behaviour change strategies used in sedentary behaviour reduction interventions among adults. *Health Psychology Review* 10, 1: 89–112. https://doi.org/10.1080/17437199.2015.1082146
34. Viswanath K Glanz K, Rimer BK. 2008. *Health behavior and health education: theory, research, and practice*.
35. Rúben Gouveia, Fábio Pereira, Evangelos Karapanos, Sean A Munson, and Marc Hassenzahl. 2016. Exploring the Design Space of Glanceable Feedback for Physical Activity Trackers. In *Proceedings of the 2016 ACM International Joint Conference on Pervasive and Ubiquitous Computing* (UbiComp '16), 144–155. https://doi.org/10.1145/2971648.2971754
36. Connor Graham, Peter Benda, Steve Howard, James Balmford, Nicole Bishop, and Ron Borland. 2006. "Heh - Keeps Me off the Smokes...": Probing Technology Support for Personal Change. In *Proceedings of the 18th Australia Conference on Computer-Human Interaction: Design: Activities, Artefacts and Environments* (OZCHI '06), 221–228. https://doi.org/10.1145/1228175.1228214
37. Kristoffer Hagen, Konstantinos Chorianopoulos, Alf Inge Wang, Letizia Jaccheri, and Stian Weie. 2016. Gameplay As Exercise. In *Proceedings of the 2016 CHI Conference Extended Abstracts on Human Factors in Computing Systems* (CHI EA '16), 1872–1878. https://doi.org/10.1145/2851581.2892515
38. Eric B. Hekler, Predrag Klasnja, Jon E. Froehlich, and Matthew P. Buman. 2013. Mind the theoretical gap. In *Proceedings of the SIGCHI Conference on Human Factors in Computing Systems - CHI '13*, 3307. https://doi.org/10.1145/2470654.2466452
39. BA Kamphorst, MCA Klein, and Arlette Van Wissen. 2014. Autonomous e-coaching in the wild: empirical validation of a model-based reasoning system. In *Proceedings of the 13th International Conference on Autonomous Agents and Multiagent Systems* (AAMAS '14), 725–732. Retrieved from http://dl.acm.org/citation.cfm?id=2615848
40. Toshikazu Kanaoka and Bilge Mutlu. 2015. Designing a Motivational Agent for Behavior Change in Physical Activity. In *Proceedings of the 33rd Annual ACM Conference Extended Abstracts on Human Factors in Computing Systems* (CHI EA '15), 1445–1450. https://doi.org/10.1145/2702613.2732924
41. Ntwa Katule, Ulrike Rivett, and Melissa Densmore. 2016. A Family Health App: Engaging Children to Manage Wellness of Adults. In *Proceedings of the 7th Annual Symposium on Computing for Development* (ACM DEV '16), 7:1--7:10. https://doi.org/10.1145/3001913.3001920
42. Sunyoung Kim, Julie A Kientz, Shwetak N Patel, and Gregory D Abowd. 2008. Are You Sleeping?: Sharing Portrayed Sleeping Status Within a Social Network. In *Proceedings of the 2008 ACM Conference on Computer Supported Cooperative Work* (CSCW '08), 619–628. https://doi.org/10.1145/1460563.1460660
43. Risa Kitagawa. 2015. Texting and Sexual Health: Experimental Evidence from an Information Intervention in Kenya. In *Proceedings of the Seventh International Conference on Information and Communication Technologies and Development* (ICTD '15), 18:1--18:10. https://doi.org/10.1145/2737856.2738032
44. Rafal Kocielnik and Gary Hsieh. 2017. Send Me a Different Message: Utilizing Cognitive Space to Create Engaging Message Triggers. *Proceedings of the 2017 ACM Conference on Computer Supported Cooperative Work and Social Computing - CSCW '17*: 2193–2207. https://doi.org/10.1145/2998181.2998324
45. Andreas Komninos, Mark D Dunlop, David Rowe, Allan Hewitt, and Steven Coull. 2015. Using Degraded Music Quality to Encourage a Health Improving Walking Pace: BeatClearWalker. In *Proceedings of the 9th International Conference on Pervasive Computing Technologies for Healthcare* (PervasiveHealth '15), 57–64. Retrieved from http://dl.acm.org/citation.cfm?id=2826165.2826174
46. Joyca Lacroix, Privender Saini, and Roger Holmes. 2008. The Relationship Between Goal Difficulty and Performance in the Context of a Physical Activity Intervention Program. In *Proceedings of the 10th International Conference on Human Computer Interaction with Mobile Devices and Services* (MobileHCI '08), 415–418. https://doi.org/10.1145/1409240.1409302
47. Jisoo Lee, Erin Walker, Winslow Burleson, Matthew Kay, Matthew Buman, and Eric B. Hekler. 2017. Self-Experimentation for Behavior Change. *Proceedings of the 2017 CHI Conference on Human Factors in Computing Systems - CHI '17*: 6837–6849. https://doi.org/10.1145/3025453.3026038
48. Jisoo Lee, Erin Walker, Winslow Burleson, Matthew Kay, Matthew Buman, and Eric B Hekler. 2017. Self-Experimentation for Behavior Change: Design and Formative Evaluation of Two Approaches. In *Proceedings of the 2017 CHI Conference on Human Factors in Computing Systems* (CHI '17), 6837–6849. https://doi.org/10.1145/3025453.3026038
49. Alessandro Liberati, Douglas G. Altman, Jennifer Tetzlaff, Cynthia Mulrow, Peter C. Gøtzsche, John P.A. Ioannidis, Mike Clarke, P. J. Devereaux, Jos Kleijnen, and David Moher. 2009. The PRISMA Statement for Reporting Systematic Reviews and Meta-Analyses of Studies That Evaluate Health Care Interventions: Explanation and Elaboration. *Annals of Internal Medicine* 151, 4: W. https://doi.org/10.7326/0003-4819-151-4-200908180-00136
50. Brian Y Lim, Aubrey Shick, Chris Harrison, and Scott E Hudson. 2011. Pediluma: Motivating Physical Activity Through Contextual Information and Social Influence. In *Proceedings of the Fifth International Conference on Tangible, Embedded, and Embodied Interaction* (TEI '11), 173–180. https://doi.org/10.1145/1935701.1935736
51. Christine Lisetti, Reza Amini, Ugan Yasavur, and Naphtali Rishe. 2013. I Can Help You Change! An Empathic Virtual Agent Delivers Behavior Change Health Interventions. *ACM Transactions on Management Information Systems* 4, 4: 1–28. https://doi.org/10.1145/2544103
52. Edwin A Locke and Gary P Latham. 2002. Building a practically useful theory of goal setting and task motivation. A 35-year odyssey. *The American psychologist* 57, 9: 705–717. https://doi.org/10.1037/0003-066X.57.9.705
53. Geke D S Ludden and Linda Meekhof. 2016. Slowing Down: Introducing Calm Persuasive Technology to Increase Wellbeing at Work. In *Proceedings of the 28th Australian Conference on Computer-Human Interaction* (OzCHI '16), 435–441. https://doi.org/10.1145/3010915.3010938
54. Elizabeth J. Lyons, Zakkoyya H. Lewis, Brian G. Mayrsohn, and Jennifer L. Rowland. 2014. Behavior change techniques implemented in electronic lifestyle activity monitors: A systematic content analysis. *Journal of Medical Internet Research* 16, 8: e192. https://doi.org/10.2196/jmir.3469
55. Andrew Macvean and Judy Robertson. 2013. Understanding exergame users' physical activity, motivation and behavior over time. In *Proceedings of the SIGCHI Conference on Human Factors in Computing Systems - CHI '13* (CHI '13), 1251. https://doi.org/10.1145/2470654.2466163
56. Laura Malinverni, Brenda López Silva, and Narc\'\is Parés. 2012. Impact of Embodied Interaction on Learning Processes: Design and Analysis of an Educational Application Based on Physical Activity. In *Proceedings of the 11th International Conference on Interaction Design and Children* (IDC '12), 60–69. https://doi.org/10.1145/2307096.2307104
57. Rebecca McCarroll, Helen Eyles, and Cliona Ni Mhurchu. 2017. Effectiveness of mobile health (mHealth) interventions for promoting healthy eating in adults: A systematic review. *Preventive Medicine* 105: 156–168. https://doi.org/10.1016/j.ypmed.2017.08.022
58. Susan Michie, Charles Abraham, Craig Whittington, John McAteer, and Sunjai Gupta. 2009. Effective techniques in healthy eating and physical activity interventions: A meta-regression. *Health Psychology* 28, 6: 690–701. https://doi.org/10.1037/a0016136
59. Susan Michie, Michelle Richardson, Marie Johnston, Charles Abraham, Jill Francis, Wendy Hardeman, Martin P. Eccles, James Cane, and Caroline E. Wood. 2013. The behavior change technique taxonomy (v1) of 93 hierarchically clustered techniques: Building an international consensus for the reporting of behavior change interventions. *Annals of Behavioral Medicine* 46, 1: 81–95. https://doi.org/10.1007/s12160-013-9486-6
60. Anouk Middelweerd, Julia S Mollee, C Natalie van der Wal, Johannes Brug, and Saskia J te Velde. 2014. Apps to promote physical activity among adults: a review and content analysis. *International Journal of Behavioral Nutrition and Physical Activity* 11, 1: 97. https://doi.org/10.1186/s12966-014-0097-9
61. Aaron S Miller, Joseph A Cafazzo, and Emily Seto. 2016. A game plan: Gamification design principles in mHealth applications for chronic disease management. *Health Informatics Journal* 22, 2: 184–193. https://doi.org/10.1177/1460458214537511
62. David C. Mohr, Stephen M. Schueller, Enid Montague, Michelle Nicole Burns, and Parisa Rashidi. 2014. The behavioral intervention technology model: An integrated conceptual and technological framework for ehealth and mhealth interventions. *Journal of Medical Internet Research* 16, 6. https://doi.org/10.2196/jmir.3077

63. Florian "Floyd" Mueller, Martin R Gibbs, and Frank Vetere. 2009. Design Influence on Social Play in Distributed Exertion Games. In *Proceedings of the SIGCHI Conference on Human Factors in Computing Systems* (CHI '09), 1539–1548. https://doi.org/10.1145/1518701.1518938
64. Sarah A. Mummah, Abby C. King, Christopher D. Gardner, and Stephen Sutton. 2016. Iterative development of Vegethon: a theory-based mobile app intervention to increase vegetable consumption. *International Journal of Behavioral Nutrition and Physical Activity* 13, 1: 90. https://doi.org/10.1186/s12966-016-0400-z
65. Sarah Ann Mummah, Maya Mathur, Abby C King, Christopher D Gardner, and Stephen Sutton. 2016. Mobile Technology for Vegetable Consumption: A Randomized Controlled Pilot Study in Overweight Adults. *JMIR mHealth and uHealth* 4, 2: e51. https://doi.org/10.2196/mhealth.5146
66. Sean A Munson, Erin Krupka, Caroline Richardson, and Paul Resnick. 2015. Effects of Public Commitments and Accountability in a Technology-Supported Physical Activity Intervention. In *Proceedings of the 33rd Annual ACM Conference on Human Factors in Computing Systems* (CHI '15), 1135–1144. https://doi.org/10.1145/2702123.2702524
67. Lennart E. Nacke, Chris Bateman, and Regan L. Mandryk. 2014. BrainHex: A neurobiological gamer typology survey. *Entertainment Computing* 5, 1: 55–62. https://doi.org/10.1016/J.ENTCOM.2013.06.002
68. Yuya Nakanishi and Yasuhiko Kitamura. 2016. Promoting Physical Activities by Massive Competition in Virtual Marathon. In *Proceedings of the Fourth International Conference on Human Agent Interaction* (HAI '16), 333–336. https://doi.org/10.1145/2974804.2980483
69. Michael I. Norton, Daniel Mochon, and Dan Ariely. 2012. The IKEA effect: When labor leads to love. *Journal of Consumer Psychology* 22, 3: 453–460. https://doi.org/10.1016/j.jcps.2011.08.002
70. Harri Oinas-kukkonen and Marja Harjumaa. 2009. Persuasive Systems Design : Key Issues , Process Model , and System Features Persuasive Systems Design : Key Issues , Process Model , and System Features. *Communications of the Association for Information Systems* 24, 1.
71. E K Olander, H Fletcher, S Williams, L Atkinson, A Turner, and D P French. 2013. What are the most effective techniques in changing obese individuals' physical activity self-efficacy and behaviour: A systematic review and meta-analysis. *The International Journal of Behavioral Nutrition and Physical Activity* 10, 29: 29. https://doi.org/10.1186/1479-5868-10-29
72. Rita Orji, Regan L. Mandryk, and Julita Vassileva. 2017. Improving the Efficacy of Games for Change Using Personalization Models. *ACM Transactions on Computer-Human Interaction* 24, 5: 1–22. https://doi.org/10.1145/3119929
73. Rita Orji and Karyn Moffatt. 2018. Persuasive technology for health and wellness: State-of-the-art and emerging trends. *Health Informatics Journal* 24, 1: 66–91. https://doi.org/10.1177/1460458216650979
74. Rita Orji, Lennart E. Nacke, and Chrysanne Di Marco. 2017. Towards Personality-driven Persuasive Health Games and Gamified Systems. In *Proceedings of the 2017 CHI Conference on Human Factors in Computing Systems - CHI '17* (CHI '17), 1015–1027. https://doi.org/10.1145/3025453.3025577
75. Jessica Pater, Shane Owens, Sarah Farmer, Elizabeth Mynatt, and Brad Fain. 2017. Addressing medication adherence technology needs in an aging population. In *Proceedings of the 11th EAI International Conference on Pervasive Computing Technologies for Healthcare - PervasiveHealth '17* (PervasiveHealth '17), 58–67. https://doi.org/10.1145/3154862.3154872
76. James O. Prochaska and Wayne F. Velicer. 1997. The transtheoretical model of health behavior change. *American Journal of Health Promotion* 12, 1: 38–48. https://doi.org/10.4278/0890-1171-12.1.38
77. Shannon Rodgers, Brittany Maloney, Bernd Ploderer, and Margot Brereton. 2016. Managing Stress, Sleep and Technologies: An Exploratory Study of Australian University Students. In *Proceedings of the 28th Australian Conference on Computer-Human Interaction* (OzCHI '16), 526–530. https://doi.org/10.1145/3010915.3010961
78. Yvonne Rogers, William R Hazlewood, Paul Marshall, Nick Dalton, and Susanna Hertrich. 2010. Ambient Influence: Can Twinkly Lights Lure and Abstract Representations Trigger Behavioral Change? In *Proceedings of the 12th ACM International Conference on Ubiquitous Computing* (UbiComp '10), 261–270. https://doi.org/10.1145/1864349.1864372
79. I M Rosenstock, V J Strecher, and M H Becker. 1988. Social learning theory and the Health Belief Model. *Health education quarterly* 15, 2: 175–83. Retrieved November 8, 2017 from http://www.ncbi.nlm.nih.gov/pubmed/3378902
80. S.Sutton. 2002. Health Beahavior - Psychosocial Theories. *University of Cambridge*: 10. https://doi.org/10.1016/B978-0-08-097086-8.14153-4
81. Akane Sano, Paul Johns, and Mary Czerwinski. 2017. Designing Opportune Stress Intervention Delivery Timing using Multi-modal Data. Retrieved March 29, 2018 from https://affect.media.mit.edu/pdfs/17.sano_etal_stressinterventiondelivery_ACII.pdf
82. Hanna Schäfer, Joachim Bachner, Sebastian Pretscher, Georg Groh, and Yolanda Demetriou. 2018. Study on Motivating Physical Activity in Children with Personalized Gamified Feedback. In *Adjunct Publication of the 26th Conference on User Modeling, Adaptation and Personalization* (UMAP '18), 221–226. https://doi.org/10.1145/3213586.3225227
83. Jaemyung Shin, Bumsoo Kang, Taiwoo Park, Jina Huh, Jinhan Kim, and Junehwa Song. 2016. BeUpright: Posture Correction Using Relational Norm Intervention. In *Proceedings of the 2016 CHI Conference on Human Factors in Computing Systems* (CHI '16), 6040–6052. https://doi.org/10.1145/2858036.2858561
84. Daniel Spelmezan, Mareike Jacobs, Anke Hilgers, and Jan Borchers. 2009. Tactile Motion Instructions for Physical Activities. In *Proceedings of the SIGCHI Conference on Human Factors in Computing Systems* (CHI '09), 2243–2252. https://doi.org/10.1145/1518701.1519044
85. Katarzyna Stawarz, Anna L Cox, and Ann Blandford. 2015. Beyond Self-Tracking and Reminders: Designing Smartphone Apps That Support Habit Formation. In *Proceedings of the 33rd Annual ACM Conference on Human Factors in Computing Systems* (CHI '15), 2653–2662. https://doi.org/10.1145/2702123.2702230
86. Tammy Toscos, Anne Faber, Shunying An, and Mona Praful Gandhi. 2006. Chick Clique: Persuasive Technology to Motivate Teenage Girls to Exercise. In *CHI '06 Extended Abstracts on Human Factors in Computing Systems* (CHI EA '06), 1873–1878. https://doi.org/10.1145/1125451.1125805
87. Roelof A.J. de Vries, Khiet P. Truong, Cristina Zaga, Jamy Li, and Vanessa Evers. 2017. A word of advice: how to tailor motivational text messages based on behavior change theory to personality and gender. *Personal and Ubiquitous Computing* 21, 4: 675–687. https://doi.org/10.1007/s00779-017-1025-1
88. Yunlong Wang, Ahmed Fadhil, Jan-Philipp Lange, and Harald Reiterer. 2019. Integrating Taxonomies Into Theory-Based Digital Health Interventions for Behavior Change: A Holistic Framework. *JMIR Research Protocols* 8, 1: e8055. https://doi.org/10.2196/resprot.8055
89. Yunlong Wang, Ulrike Pfeil, and Harald Reiterer. 2016. Supporting Self-Assembly. In *Proceedings of the 2016 ACM Workshop on Multimedia for Personal Health and Health Care - MMHealth '16*, 19–22. https://doi.org/10.1145/2985766.2985775
90. Yunlong Wang, Lingdan Wu, Jan-Philipp Lange, Ahmed Fadhil, and Harald Reiterer. 2017. Persuasive Technology in Reducing Prolonged Sedentary Behavior at Work: A Systematic Review. *arXiv.org*. Retrieved March 26, 2018 from https://arxiv.org/ftp/arxiv/papers/1712/1712.02547.pdf
91. Yunlong Wang, Lingdan Wu, Jan-Philipp Lange, Ahmed Fadhil, and Harald Reiterer. 2018. Persuasive Technology in Reducing Prolonged Sedentary Behavior at Work: A Systematic Review. *Smart Health*. https://doi.org/10.1016/j.smhl.2018.05.002
92. Isaac Wiafe, Keiichi Nakata, and Stephen Gulliver. 2014. Categorizing users in behavior change support systems based on cognitive dissonance. *Personal and Ubiquitous Computing* 18, 7: 1677–1687. https://doi.org/10.1007/s00779-014-0782-3
93. Emi Yuda, Akira Kurata, Yutaka Yoshida, and Junichiro Hayano. 2017. Life Style Modification by Peer Monitoring of Physical Activity. In *Proceedings of the 14th EAI International Conference on Mobile and Ubiquitous Systems: Computing, Networking and Services* (MobiQuitous 2017), 520–522. https://doi.org/10.1145/3144457.3145494
94. Tae-Jung Yun and Rosa I. Arriaga. 2016. SMS is my BFF: positive impact of a texting intervention on low-income children with asthma. *Proceedings of the 10th EAI International Conference on Pervasive Computing Technologies for Healthcare*: 53–60. Retrieved September 17, 2018 from https://dl.acm.org/citation.cfm?id=3021328
95. Oren Zuckerman and Ayelet Gal-Oz. 2014. Deconstructing gamification: evaluating the effectiveness of continuous measurement, virtual rewards, and social comparison for promoting physical activity. *Personal and Ubiquitous Computing* 18, 7: 1705–1719. https://doi.org/10.1007/s00779-014-0783-2